\documentclass[conference]{IEEEtran}
\usepackage{algorithm,algorithmic}
\makeatletter
\newcommand\fs@betterruled{%
  \def\@fs@cfont{\bfseries}\let\@fs@capt\floatc@ruled
  \def\@fs@pre{\vspace*{5pt}\hrule height.8pt depth0pt \kern2pt}%
  \def\@fs@post{\kern2pt\hrule\relax}%
  \def\@fs@mid{\kern2pt\hrule\kern2pt}%
  \let\@fs@iftopcapt\iftrue}
\floatstyle{betterruled}
\restylefloat{algorithm}

\makeatother
\newcommand{\squeezeup}{\vspace{-4mm}}
\newcommand{\squeezeupan}{\vspace{-2mm}}
\newcommand{\squeezeupann}{\vspace{-1mm}}

\usepackage{cite}
\usepackage{amsmath,amssymb,amsfonts}
\usepackage{algorithmic}
\usepackage{graphicx}
\usepackage{textcomp}
\usepackage{comment}
\usepackage{enumitem}
\usepackage{xcolor}
\def\BibTeX{{\rm B\kern-.05em{\sc i\kern-.025em b}\kern-.08em
    T\kern-.1667em\lower.7ex\hbox{E}\kern-.125emX}}
\renewcommand{\a}{\mathbf{a}}

\newcommand{\g}{\mathbf{g}}

\newcommand{\n}{\mathbf{n}}

\renewcommand{\r}{\mathbf{r}}
\newcommand{\s}{\mathbf{s}}
\renewcommand{\t}{\mathbf{t}}

\renewcommand{\v}{\mathbf{v}}
\newcommand{\w}{\mathbf{w}}
\newcommand{\x}{\mathbf{x}}
\newcommand{\y}{\mathbf{y}}
\newcommand{\z}{\mathbf{z}}

\newcommand{\B}{\mathbf{B}}
\newcommand{\C}{\mathbf{C}}
\newcommand{\D}{\mathbf{D}}

\newcommand{\F}{\mathbf{F}}
\newcommand{\G}{\mathbf{G}}
\renewcommand{\H}{\mathbf{H}}
\newcommand{\I}{\mathbf{I}}

\newcommand{\R}{\mathbf{R}}

\newcommand{\U}{\mathbf{U}}
\newcommand{\V}{\mathbf{V}}
\newcommand{\W}{\mathbf{W}}

\newcommand{\Compl}{\mbox{$\mathbb{C}$}}

\newcommand{\argmax}{\operatornamewithlimits{argmax}}

\begin{document}
\title{Integrated Sensing and Communication  with Millimeter Wave Full Duplex Hybrid Beamforming}

\author{\IEEEauthorblockN{Md Atiqul Islam\IEEEauthorrefmark{2}, George C. Alexandropoulos\IEEEauthorrefmark{4}, and Besma Smida\IEEEauthorrefmark{2}}
\IEEEauthorblockA{{\IEEEauthorrefmark{2}Department of Electrical and Computer Engineering, University of Illinois at Chicago, USA}\\
\IEEEauthorrefmark{4}Department of Informatics and Telecommunications, National and Kapodistrian University of Athens, Greece\\
emails: \{mislam23,smida\}@uic.edu, alexandg@di.uoa.gr
}}
\maketitle
\begin{abstract}
Integrated Sensing and Communication (ISAC) has attracted substantial attraction in recent years for spectral efficiency improvement, enabling hardware and spectrum sharing for simultaneous sensing and signaling operations. In-band Full Duplex (FD) is being considered as a key enabling technology for ISAC applications due to its simultaneous transmission and reception capability. In this paper, we present an FD-based ISAC system operating at millimeter Wave (mmWave) frequencies, where a massive Multiple-Input Multiple-Output (MIMO) Base Station (BS) node employing hybrid Analog and Digital (A/D) beamforming is communicating with a DownLink (DL) multi-antenna user and the same waveform is utilized at the BS receiver for sensing the radar targets in its coverage environment. We develop a sensing algorithm that is capable of estimating Direction of Arrival (DoA), range, and relative velocity of the radar targets. A joint optimization framework for designing the A/D transmit and receive beamformers as well as the Self-Interference (SI) cancellation is presented with the objective to maximize the achievable DL rate and the accuracy of the radar target sensing performance. Our simulation results, considering fifth Generation (5G) Orthogonal Frequency Division Multiplexing (OFDM) waveforms, verify our approach's high precision in estimating DoA, range, and velocity of multiple radar targets, while maximizing the DL communication rate.
\end{abstract}


\section{Introduction}
Integrated Sensing and Communication (ISAC) is an emerging concept for future wireless networks, where the previously competing sensing and communication operations are jointly optimized in the same hardware platform using a unified signal processing framework \cite{liu2020joint,HRIS,alexandropoulos2021hybrid_all,mishra2019toward,paul2016survey}. Recently, Full Duplex (FD) massive Multiple-Input Multiple-Output (MIMO) communications have been considered a key enabler for ISAC applications due to their simultaneous UpLink (UL) and DownLink (DL) transmission capability within the entire frequency band \cite{B:Full-Duplex,alexandropoulos2020full,islam2021direction,Islam_2020_Sim_Multi}. Furthermore, FD massive MIMO ISAC applications at millimeter Wave (mmWave) frequencies have the potential to provide high capacity communication links while simultaneously achieving high-resolution sensing, e.g., Direction of Arrival (DoA), range, and relative speed of radar targets/scatterers.

The performance of the FD ISAC systems relies on the in-band Self-Interference (SI) signal suppression capability that stems from the Transmitter (TX) to the Receiver (RX) side during FD operation. Recently in \cite{islam2021direction,xiao2017full_all,Vishwanath_2020,alexandropoulos2020full}, SI cancellation was achieved for the FD massive MIMO systems operating at mmWave, utilizing a combination of propagation domain isolation, analog domain suppression, and digital SI cancellation techniques. To alleviate the hardware cost in mmWave massive MIMO transceivers, hybrid Analog and Digital (A/D) beamformers are usually employed, where large-scale antenna arrays are usually connected to a small number of Radio Frequency (RF) chains via analog preprocessing networks comprised of phase shifters \cite{venkateswaran2010analog}. Such systems require appropriate beam selection for analog TX/RX beamformers, chosen from predefined codebooks, to maximize DL rate and sensing accuracy. Moreover, in the envisioned FD ISAC with massive MIMO radios, the transmit waveform will be utilized for both DL data transmission and sensing of the radar targets \cite{liu2020joint}. Therefore, a joint design of the A/D beamformers and SI cancellation along with sensing techniques is required for maximizing the performance of FD ISAC systems.

Very recently in \cite{barneto2019full,liyanaarachchi2021optimized}, joint radar communication and sensing frameworks, leveraging FD operation, were considered for single-antenna systems, where both communication and radar waveforms were optimized for sensing performance. In \cite{barneto2020beamforming,liyanaarachchi2021joint}, the FD ISAC operation was proposed for mmWave frequency bands considering a massive MIMO FD Base Station (BS), where the signal power is maximized in the radar target direction, while maintaining a threshold DL rate performance. Although the considered FD ISAC approach estimated the DoA two radar targets, the range was only calculated for one target due to its disassociated DoA and range estimation technique. 
\begin{figure*}[!tpb]
	\begin{center}
	\includegraphics[width=0.95\linewidth]{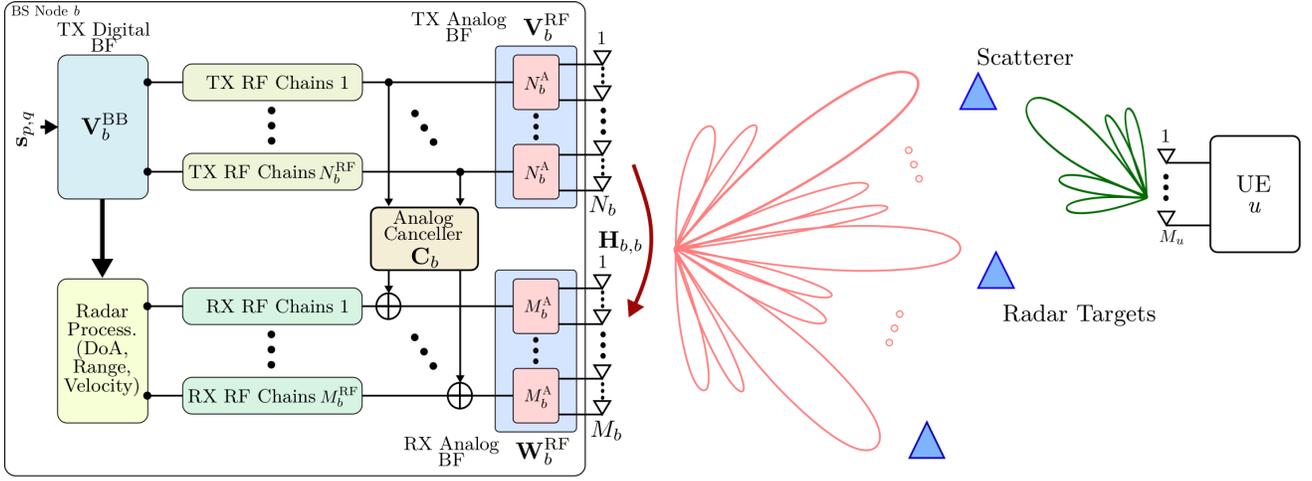}
	\caption{The considered FD massive MIMO ISAC system: the FD hybrid A/D beamforming Base Station (BS) communicates in the downlink with a mobile single-antenna half-duplex User Equipment (UE), while its reflected transmitted signals from radar targets/scatterers in the environment are received via hybrid combining and processed for DoA, range, and relative velocity estimations of the radar targets/scatterers.}
	\label{fig: FD_ISAC}
	\end{center}
	\squeezeup
	\squeezeupann
\end{figure*}

In this paper, we present a novel FD massive MIMO ISAC system operating at mmWave frequencies and realizing hybrid A/D beamforming, where Orthogonal Frequency Division Multiplexing (OFDM) waveforms are utilized for both DL communication and radar target sensing. Unlike state-of-the-art works, we devise an ISAC optimization framework that is capable of estimating the DoA, range, and relative velocity of multiple radar targets, while maximizing the DL communication rate. Our numerical results, considering the fifth Generation (5G) New Radio (NR) OFDM waveform, verify the high sensing accuracy and the increased communication rate of the proposed optimization design.  

\section{System and Signal Models}
We consider an FD ISAC system comprising of an FD mmWave massive MIMO Base Station (BS) node $b$ equipped with $N_b$ TX and $M_b$ RX antenna elements communicating in the DL direction with an RX User Equipment (UE) node $u$ with $M_u$ antenna elements, while the reflected DL signal is utilized to detect radar targets/scatterers randomly distributed within the communication/sensing environment at the RX of BS node $b$, as depicted in Fig. \ref{fig: FD_ISAC}. To reduce the number of RF chains and phase shifters, the BS node $b$ employs a partially-connected Hybrid BeamForming (HBF) structure with A/D TX and RX beamformers, where each of the
$N_{b}^{\rm RF}$ and $M_{b}^{\rm RF}$ TX/RX RF chains is connected to a Uniform Linear Arrays (ULA) of $N_{b}^{\rm A}$ and $M_{b}^{\rm A}$ antenna elements, respectively, via phase shifters. Therefore, it holds $N_b = N_{b}^{\rm A}N_{b}^{\rm RF}$ and $M_b = M_{b}^{\rm A}M_{b}^{\rm RF}$ for total number of TX and RX antennas, respectively, at the BS node $b$. Since the size of the antenna array at the UE node $u$ is typically much smaller than at the FD massive MIMO BS $b$, we assume that the UE adopts a fully digital beamforming structure.

It is assumed that BS node $b$ transmits mmWave OFDM waveforms in the DL direction containing $Q$ OFDM symbols with $P$ active subcarriers. In the BaseBand (BB), the unit power frequency-domain symbol vector $\s_{p,q}\in\Compl^{d_b\times1}$ at the $p$th subcarrier of $q$th OFDM symbol is precoded using digital beamforming matrix $\V_b^{\mathrm{BB}}\in\Compl^{N_{b}^{\rm RF}\times d_b}$, where $d_b\leq\min\{N_{b}^{\mathrm{RF}},M_u\}$. Following the BB precoder, the DL signal is processed by the analog beamformer $\V_b^{\mathrm{RF}}\in\Compl^{N_{b}\times N_{b}^{\rm RF}}$ containing the configurations of the phase shifters as follows:
\begin{equation}\label{eq:TX_analog_precoder}
\mathbf{V}_b^{\mathrm{RF}} \triangleq \left[ \begin{matrix}
\mathbf{v}_1  &  \mathbf{0}_{N_b^{{\rm A}}\times1}    &  \cdots    &  \mathbf{0}_{N_b^{{\rm A}}\times1}  \\
\mathbf{0}_{N_b^{{\rm A}}\times1}    &  \mathbf{v}_2  &  \cdots    &  \mathbf{0}_{N_b^{{\rm A}}\times1}  \\
\vdots   &  \vdots   &  \ddots    &	 \vdots  \\
\mathbf{0}_{N_b^{{\rm A}}\times1}    &  \mathbf{0}_{N_b^{{\rm A}}\times1}    &  \cdots    &  \mathbf{v}_{N_b^{({\rm RF})}}
\end{matrix}
\right].
\end{equation}
The elements of each $\v_i$ are assumed to have constant magnitude, i$.$e$.$, $|[\mathbf{v}_i]_{n}|^2=1/N_b^{{\rm A}}$ $\forall$$n=1,2,\ldots,N_b^{{\rm A}}$. We also assume that $\mathbf{v}_i\in\mathbb{F}_{\rm TX}$ $\forall$$i=1,2,\ldots,N_{b}^{\mathrm{RF}}$, which means that all analog TX precoding vectors belong in a predefined beam codebook $\mathbb{F}_{\rm TX}$ including ${\rm card}(\mathbb{F}_{\rm TX})$ distinct vectors (or analog beams). Applying both A/D beamforming, the TX frequency-domain
symbols at the antenna elements are expressed as
\begin{equation}\label{eq:TX_Signal}
\x_{p,q} \triangleq \V_b^{\mathrm{RF}}\V_b^{\mathrm{BB}}\s_{p,q},
\end{equation}
where $\x_{p,q}\in\Compl^{N_b\times 1}$. The signal transmissions at node $b$ is power limited according to $\mathbb{E}\{\|\V_b^{\mathrm{RF}}\V_b^{\mathrm{BB}}\s_{p,q}\|^2\}\leq {\rm P}_b$.

\subsection{Radar Signal Reception Model}
For sensing operation, we consider that a collection of $K$ radar targets/scatterers are randomly distributed within the communication/sensing environment and are to be detected by the BS node $b$. All the $K$ targets reflect DL signal back to the RX of the BS node, while only a subset of them ($L$ out of $K$) contributes to the DL communication scattering paths between the BS node $b$ and the RX UE node $u$. The purpose of the sensing operation is to estimate the DoA, range, and relative velocity of each radar targets.

We consider that the DoAs of the $K$ targets are defined as $\Theta =[\theta_1,\theta_2,\cdots,\theta_{K}]$, while the distance and the relative speed of $k$th target correspond to a delay $\tau_k$ and a Doppler shift $f_{D,k}$, respectively. Enabled by FD, the received signal $\y_{p,q}\in\mathbb{C}^{M_{b}\times 1}$ at the RX of the BS node $b$ combining the SI and the reflected signal by the Radar targets is expressed as
\begin{equation}\label{eq:Radar_Signal}
    \begin{split}
        \y_{p,q} \triangleq & \sum\limits_{k=1}^{K} \alpha_k e^{j2\pi(qT_s f_{D,k} - p\tau_k \Delta f)} \a_{M_b}(\theta_k)\a_{N_b}^{\rm H}(\theta_k) \x_{p,q}\\
        &+ \H_{b,b} \x_{p,q} + \n_{p,q},
    \end{split}
\end{equation}
where $\alpha_k\in\Compl$ and $\n_{p,q}\sim \mathcal{CN}(0,\sigma_b^2\I_{M_b})$ represent the reflection coefficient of the $k$th radar target and the receiver noise floor, respectively. Here, $\Delta f$ and $T_s \triangleq 1/\Delta f + T_{cp}$ denote the subcarrier spacing and the total OFDM symbol duration (including the cyclic prefix). The propagation delay causes the phase shift $e^{-j2\pi p\tau_k\Delta f}$ across subcarriers, while the Doppler shift contributes a row-wise oscillation across different OFDM symbols \cite{braun2010maximum}. Considering ULA, the steering vector $\a_{N_b}(\theta)$ for $N_b$ antenna elements and any DoA $\theta$ is formulated as \cite{gonzalez2018channel}
\begin{equation}
    \begin{split}
        \a_{M_b}(\theta)\! \triangleq \! \frac{1}{\sqrt{N_b}} \big[1,e^{j\frac{2\pi}{\lambda}d \sin(\theta)},\ldots,e^{j\frac{2\pi}{\lambda}(N_b-1)d \sin(\theta)}]^{\rm T},
    \end{split}
\end{equation}
where $\lambda$ is the propagation signal wavelength and $d$ denotes the distance between adjacent antenna elements.
Here, $\mathbf{H}_{b,b}\in\Compl^{M_b \times N_b}$ is the SI channel path at the BS node $b$, which is modeled as a Rician fading mmWave channel as \cite[eq. (9)]{satyanarayana2018hybrid}. 

The received signal at the node $b$ RX is first processed by the RF combiner $\W_b^{\mathrm{RF}}\in\Compl^{M_b \times M_{b}^{\rm RF}}$, where the structure of the combiner is formulated similarly as \eqref{eq:TX_analog_precoder}. Here, the analog RX combining vectors belong in a predefined beam codebook $\mathbb{F}_{\rm RX}$ including ${\rm card}(\mathbb{F}_{\rm RX})$ distinct vectors. After RF combination and A/D SI cancellation, the received signal is expressed as
\begin{align}
        \nonumber\widetilde{\y}_{p,q}\! \triangleq &  ( \W_{b}^{\rm RF})^{\rm H}\!\sum\limits_{k=1}^{K} \!\!\alpha_k e^{j2\pi(qT_s f_{D,k} - p\tau_k \Delta f)} \a_{M_b}\!(\theta_k)\a_{N_b}^{\rm H}\!(\theta_k) \x_{p,q}\\
        &+  (\widetilde{\H}_{b,b} \!+\! \C_{b} \!+\! \D_{b})\V_{b}^{\rm BB}\s_{p,q}\! + \!(\W_{b}^{\rm RF})^{\rm H}\n_{p,q},
\end{align}
where $\widetilde{\H}_{b,b}\triangleq(\W_{b}^{\rm RF})^{\rm H}\H_{b,b}\V_{b}^{\rm RF}$ is the effective SI channel after analog TX/RX beamforming. Here, $\C_{b}$ and $\D_{b}$ represent the analog and digital SI cancellation, respectively, that are designed following the structure presented in \cite{alexandropoulos2020full}.

\subsection{DL Signal Reception Model}
As mentioned above, $L$ out of $K$ scatterers contributes to the DL channel $\H_{u,b}\in\Compl^{M_u \times N_b}$. As the principal focus of the paper is to estimate Radar target parameters, we ignore the delay and Doppler shift parameters for the DL channel. Now, the received DL signal vector $\r_{p,q}\in\Compl^{M_u \times d_b}$ at the UE RX is expressed as
\begin{equation}
    \begin{split}
        \r_{p,q} &\triangleq \W_{u}^{\rm H}\Big(\sum\limits_{\ell=1}^{L} \beta_{\ell}  \mathbf{a}_{M_u}(\theta_{\ell})\mathbf{a}_{N_b}^{\rm H}(\theta_{\ell})\x_{p,q} + \z_{p,q}\Big)\\
        &=\W_{u}^{\rm H}\Big(\mathbf{H}_{u,b}\x_{p,q} + \z_{p,q}\Big),
    \end{split}
\end{equation}
where $\beta_{\ell}\in\Compl$ and $\z_{p,q}\sim \mathcal{CN}(0,\sigma_u^2\I_{M_u})$ represent the complex reflection coefficient of $\ell$th scatter path and the noise floor at RX node $u$, respectively.

The achievable DL rate of the FD ISAC system can be expressed as
\begin{align}
        \nonumber\mathcal{R}_{\rm DL} \triangleq \log_2\Big({\rm det}\Big(&\I_{d_b}+\W_{u}^{\rm H}\mathbf{H}_{u,b}\mathbf{V}_{b}^{\rm RF}\mathbf{V}_{b}^{\rm BB}
        (\mathbf{V}_{b}^{\rm RF}\mathbf{V}_{b}^{\rm BB})^{\rm H}\\
        &\times\mathbf{H}_{u,b}^{\rm H}\W_{u}(\W_{u}^{\rm H}\W_{u}\sigma_u^2)^{-1}\Big)\Big),
\end{align}

\section{DoA, Delay, and Doppler Shift Estimation}
In this section, we present the estimations of the DoA, delay, and Doppler shift parameters of the Radar targets, which are realized by the BS's RX using the reflected signals.

\subsection{Radar Target DoA Estimation}
For DoA estimation, we deploy the MUltiple SIgnal Classification (MUSIC) algorithm; other DoA estimation techniques can be used as well \cite{krim1996two}. 
First, we estimate the covariance matrix of the radar target reflected signal. Across all subcarriers and OFDM symbols of the communication slot, the covariance matrix $\R_{b}\in\Compl^{M_{b}^{\rm RF}\times M_{b}^{\rm RF}}$ can be estimated as
\begin{equation}\label{eq: sampl_cov}
    \begin{split}
        {\R}_b\! \triangleq\!\mathbb{E}\{\widetilde{\y}_{p,q}\widetilde{\y}_{p,q}^{\rm H}\} ,\, \widehat{\R}_b\triangleq\frac{1}{PQ}  \sum\limits_{q=0}^{Q-1}\sum\limits_{p=0}^{P-1}\widetilde{\y}_{p,q}\widetilde{\y}_{p,q}^{\rm H}.
    \end{split}
\end{equation}
By taking the eigenvalue decomposition of the estimated sample covariance matrix $\widehat{\R}_b$, it is deduced that:
\begin{equation}\label{eq: cov_eig}
    \begin{split}
        \widehat{\R}_b \triangleq \U {\rm diag}\{\eta_1,\eta_2,\ldots,\eta_{M_b}\}\U^{\rm H},
    \end{split}
\end{equation}
where $\eta_1\geq\eta_2\geq\ldots\geq\eta_{M_{b}^{\rm RF}}$ are the eigenvalues of $\widehat{\R}_b$ and $\U\in\mathbb{C}^{M_{b}^{\rm RF}\times M_{b}^{\rm RF}}$ contains their corresponding eigenvectors. Since we are interested in estimating the DoAs of $K$ radar targets, the matrix $\U$ can be partitioned as $\U=[\U_s|\U_n]$, where the columns in $\U_{n}\in\mathbb{C}^{M_{b}^{\rm RF}\times M_{b}^{\rm RF}-K}$ are the eigenvectors spanning the noise subspace and $\U_s\in\mathbb{C}^{M_{b}^{\rm RF}\times K}$ contains the signal space eigenvectors. The MUSIC spectrum for the considered HBF architecture can be thus formulated as:
\begin{equation}\label{eq: spectral_peak}
    \begin{split}
        S(\theta) \triangleq \left(\a_{M_b}^{\rm H}(\theta)\W_{b}^{\rm RF}\U_{n}\U_{n}^{\rm H}(\W_{b}^{\rm RF})^{\rm H}\a_{M_b}(\theta)\right)^{-1},
    \end{split}
\end{equation}
whose $K$ peaks correspond to the $K$ estimated DoAs $\widehat{\theta_k},\forall k$.

\begin{algorithm}[!t]
    \caption{Delay and Doppler Shift Estimation}
    \label{alg:the_est}
    \begin{algorithmic}[1]
        \renewcommand{\algorithmicrequire}{\textbf{Input:}}
        \renewcommand{\algorithmicensure}{\textbf{Output:}}
        \REQUIRE $\x_{p,q}$, $\widetilde{\y}_{p,q}$,$\forall p,q$, $\W_{b}^{\rm RF}$, and $\widehat{\theta}_k\,\,\forall k$. 
        \ENSURE $\widehat{\tau}_k, \widehat{f}_{D,k}$, $\forall k$.
        \STATE Set $n = 0,\cdots,P-1$ and $m=-\frac{Q}{2},\cdots,\frac{Q}{2}-1$.
        \FOR{$k= 1,2,\ldots,K$}
            \STATE Set $\g_{p,q} \triangleq \mathbf{a}_{M_b}(\widehat{\theta}_k)\mathbf{a}_{N_b}^{\rm H}(\widehat{\theta}_k)\x_{p,q}$.
            \STATE Set $z_{p,q} \triangleq \frac{1}{M_{b}}\sum\limits_{i=1}^{M_{b}}[\W_{b}^{\rm RF}\widetilde{\mathbf{y}}_{p,q}]_i/[\g_{p,q}]_i,$\,\,$\forall p,q$.
            \STATE Set $A(n,m) \triangleq \sum\limits_{p=0}^{P-1}\left(\sum\limits_{q=0}^{Q-1}
            z_{p,q} e^{-j2\pi\frac{qm}{Q}}\right)e^{j2\pi\frac{pn}{P}}$.
            \STATE Set $({n}^*,{m}^*) = \underset{n,m}{\text{arg max}}\quad |A(n,m)|^2$.
            \STATE Set the estimated delay $\widehat{\tau}_k = \frac{{n}^*}{P\Delta f}$.
            \STATE Set the Doppler frequency of $k$th target $\widehat{f}_{D,k} = \frac{{m}^*}{QT_s}$.
        \ENDFOR
    \end{algorithmic}
\end{algorithm}
\subsection{Delay and Doppler Shift Estimation}
The next step is to estimate the delay and Doppler shift parameters associated with the $K$ estimated DoAs. Using the estimate DoA $\widehat{\theta}_k$ and the known transmit signal $\x_{p,q}$, we formulate a reference signal in the DoA direction as
\begin{equation}
    \begin{split}
        \g_{p,q} \triangleq \mathbf{a}_{M_b}(\widehat{\theta}_k)\mathbf{a}_{N_b}^{\rm H}(\widehat{\theta}_k)\x_{p,q}.
    \end{split}
\end{equation}
Now, we utilize the received signal $\widetilde{\y}_{p,q}$ to derive the quotient averaged across all RX antennas that includes the effect of delay and Doppler shift in the direction of $\widehat{\theta}_k$ as
\begin{equation}
    \begin{split}
        z_{p,q} \triangleq \frac{1}{M_{b}}\sum\limits_{i=1}^{M_{b}}[\W_{b}^{\rm RF}\widetilde{\mathbf{y}}_{p,q}]_i/[\g_{p,q}]_i,\,\,\forall p,q.
    \end{split}
\end{equation}
To estimate $\widehat{\tau}_k$ and $ \widehat{f}_{D,k}$, we formulate the likelihood function:
\begin{equation}
    \begin{split}
        A(n,m) \triangleq \sum\limits_{p=0}^{P-1}\left(\sum\limits_{q=0}^{Q-1}
            z_{p,q} e^{-j2\pi\frac{qm}{Q}}\right)e^{j2\pi\frac{pn}{P}},
    \end{split}
\end{equation}
where $n = 0,\cdots,P-1$ and $m=-Q/2,\cdots,Q/2-1$. Finally, we find the best quantized delay and Doppler shift that maximizes the likelihood function norm. The delay and Doppler shift estimation procedure is provided in Algorithm \ref{alg:the_est}.

\section{Proposed ISAC Optimization Framework}
In this section, we present a joint optimization framework deriving A/D beamformers and SI cancellation matrices with the objective to maximize the DL rate and the radar estimation accuracy.

We consider a time division duplexing communication protocol, where the DoAs estimated in one communication time slot is utilized to derive the beamformers and SI cancellation matrices for the successive slot. 
To optimize the DL rate and radar target parameter estimation accuracy, we propose to maximize the SNR in both the DL and radar target direction. Given the estimated DoAs $\widehat{\theta}_k,\,\forall k$, the SNR in the direction of all radar targets can be written as
\begin{equation}
    \begin{split}
        \widehat{ {\Gamma}}_{\rm Radar} \triangleq \Big\|(\mathbf{W}_{b}^{\rm RF})^{\rm H}\sum\limits_{k=1}^{K} \mathbf{a}_{M_b}(\widehat{\theta}_k)\mathbf{a}_{N_b}^{\rm H}(\widehat{\theta}_k)\mathbf{V}_{b}^{\rm RF}\mathbf{V}_{b}^{\rm BB}\Big\|^2 {\Sigma}_b^{-1},
    \end{split}
\end{equation}
where $ {\Sigma}_b = \|(\widehat{\widetilde{\H}}_{b,b} + \C_{b}+\D_{b})\mathbf{V}_{b}^{\rm BB}\|^2 + \left\|\mathbf{W}_{b}^{\rm RF}\right\|^2\sigma_b^2$ is the interference plus noise covariance matrix at the RX of node $b$. Similarly, the estimated DL SNR is expressed as
\begin{equation}
    \begin{split}
        \widehat{ {\Gamma}}_{\rm DL} = \Big\|\mathbf{W}_{u}^{\rm H}\sum\limits_{\ell=1}^{L} \mathbf{a}_{M_u}(\widehat{\theta}_{\ell})\mathbf{a}_{N_b}^{\rm H}(\widehat{\theta}_{\ell})\mathbf{V}_{b}^{\rm RF}\mathbf{V}_{b}^{\rm BB}\Big\|^2 {\Sigma}_u^{-1},
    \end{split}
\end{equation}
where $ {\Sigma}_u =  \left\|\mathbf{W}_{u}\right\|^2\sigma_u^2$ is the noise covariance matrix at the RX node $u$.

The optimization problem to maximize the Radar target and DL SNR can be written as
\begin{align}\label{eq: optimization_eq}
        \mathcal{OP}&:\nonumber\underset{\substack{\mathbf{V}_{b}^{\rm RF},\mathbf{V}_{b}^{\rm BB},\mathbf{W}_{b}^{\rm RF}\\ \C_b,\D_b,\mathbf{W}_{u}}}{\max} \quad \widehat{ {\Gamma}}_{\rm Radar} + \widehat{ {\Gamma}}_{\rm DL}\\
        &\text{\text{s}.\text{t}.}\quad
        \left\|[(\widehat{\widetilde{\H}}_{b,b}\! +\! \C_{b})\V_{b}^{\rm BB}]_{(j,:)}\right\|^2 \!\!\!\leq \! \lambda_b , \!\forall\! j\! = 1, \ldots, M_{b}^{\rm RF},\nonumber\\
        &\quad\quad\mathbb{E}\{\|\mathbf{V}_{b}^{\rm RF}\mathbf{V}_{b}^{\rm BB}\|^2\}\leq {\rm P}_b,\\
        &\quad\quad\mathbf{w}_j\in\mathbb{F}_{\rm RX}\,\,\forall j\,\,{\rm and}\,\, \mathbf{v}_n\in\mathbb{F}_{\rm TX}\,\,\forall n=1,2,\ldots,N_b^{({\rm RF})}\nonumber
\end{align}

\begin{algorithm}[!t]
    \caption{FD ISAC Optimization}
    \label{alg:the_opt}
    \begin{algorithmic}[1]
        \renewcommand{\algorithmicrequire}{\textbf{Input:}}
        \renewcommand{\algorithmicensure}{\textbf{Output:}}
        \REQUIRE $\widehat{\H}_{b,b}$, $N$, ${\rm P}_b$ and $\widehat{\theta}_k\,\,\forall k$. 
        \ENSURE $\mathbf{V}_{b}^{\rm RF},\mathbf{V}_{b}^{\rm BB}, \mathbf{W}_{b}^{\rm RF},\C_b,\D_b$, and $\mathbf{W}_{u}$.
        \STATE Set $\widehat{\H}_{\rm R} \triangleq \sum\limits_{k=1}^{K} \mathbf{a}_{M_b}(\widehat{\theta}_k)\mathbf{a}_{N_b}^{\rm H}(\widehat{\theta}_k)$.
        \STATE Set $\widehat{\H}_{\rm u,b} \triangleq \sum\limits_{\ell=1}^{L} \mathbf{a}_{M_u}(\widehat{\theta}_{\ell})\mathbf{a}_{N_b}^{\rm H}(\widehat{\theta}_{\ell})$.
        \STATE Set $\mathbf{W}_{u}$ as the $d_b$ left-singular vectors of $\widehat{\H}_{\rm u,b}$ corresponding to the singular values in descending order.
        \STATE Set $\mathbf{V}_{b}^{\rm RF} \triangleq \underset{\v_j\in\mathbb{F}_{\rm TX}}{\text{arg max}}\quad \|\widehat{\H}_{\rm R} \mathbf{V}_{b}^{\rm RF}\|^2$.
        \STATE Set $\mathbf{W}_{b}^{\rm RF} \triangleq \underset{\w_j\in\mathbb{F}_{\rm TX}}{\text{arg max}}\quad \frac{\|(\mathbf{W}_{b}^{\rm RF})^{\rm H}\widehat{\H}_{\rm R} \mathbf{V}_{b}^{\rm RF}\|^2}{\|(\mathbf{W}_{b}^{\rm RF})^{\rm H}\widehat{\H}_{\rm b,b} \mathbf{V}_{b}^{\rm RF}\|^2}$.
        \STATE Set $\widehat{\widetilde{\H}}_{b,b} = (\mathbf{W}_{b}^{\rm RF})^{\rm H}\widehat{\H}_{b,b}\mathbf{V}_{b}^{\rm RF}$ and $\widehat{\widetilde{\H}}_{u,b} = \widehat{\H}_{u,b}\mathbf{V}_{b}^{\rm RF}$.
        \STATE Set $\C_b = \begin{bmatrix}-[\widehat{\widetilde{\H}}_{b,b}]_{:,1:\frac{N}{{M}_{b}^{\rm RF}}} & \mathbf{0}_{:,({N}_{b}^{\rm RF}-\frac{N}{{M}_{b}^{\rm RF}}:{N}_{b}^{\rm RF})}
                        \end{bmatrix}$ and $\D_b = -(\widehat{\widetilde{\H}}_{b,b}+C_b)$.
        \STATE Obtain $\B$ with the $N_b^{\rm RF}$ right-singular vectors of $(\widehat{\widetilde{\H}}_{b,b}+\C_b)$ corresponding to the singular values in descending order.
        \FOR{$\alpha={N}_{b}^{\rm RF},N_b^{\rm RF}-1,\ldots,2$}
    		\STATE Set $\F=[\B]_{(:,N_b^{\rm RF}-\alpha+1:N_b^{\rm RF})}$.
    		\STATE Set $\G$ as the optimum precoding for the effective DL MIMO channel $\widehat{\widetilde{\H}}_{u,b}\F$ given ${\rm P}_b$.
    		\IF{$\left\|[(\widehat{\widetilde{\H}}_{b,b}\! +\! \C_{b})\F\G]_{(j,:)}\right\|^2 \!\!\!\leq \! \lambda_b , \!\forall\! j\! = 1, \ldots, M_{b}^{\rm RF}$,}
    			 \STATE Output $\V_b^{\mathrm{BB}}=\F\G$ and stop the algorithm.
    	    \ELSE
            			 \STATE Output that the $\C_b$ realization does not meet 
            			 the residual SI constraint.
    		\ENDIF
    	\ENDFOR
    \end{algorithmic}
\end{algorithm}

The optimization problem in \eqref{eq: optimization_eq} is a non-convex problem with coupling variables, hence, quite difficult to tackle. In this work, we solve it suboptimally using alternating optimization, leaving other possibilities for future work.

First, using the estimated DoAs, we formulate a virtual channel of in the radar target direction as $\widehat{\H}_{\rm R} \triangleq \sum\limits_{k=1}^{K} \mathbf{a}_{M_b}(\widehat{\theta}_k)\mathbf{a}_{N_b}^{\rm H}(\widehat{\theta}_k)$. Now to maximize the radar SNR, we find the TX analog beams solving the following suboptimization problem:
\begin{align}\label{eq: sub_opt1}
        \mathcal{OP}1:&\quad\mathbf{V}_{b}^{\rm RF} \triangleq \underset{\v_j\in\mathbb{F}_{\rm TX}}{\argmax}\quad \|\widehat{\H}_{\rm R} \mathbf{V}_{b}^{\rm RF}\|^2
\end{align}
The solution of $\mathcal{OP}1$ requires a simple search through the beam codebook $\mathbb{F}_{\rm TX}$. Using the TX analog beamformer $\mathbf{V}_{b}^{\rm RF}$, we derive the analog combiner solving the suboptimization problem $\mathcal{OP}1$, where we simultaneously maximize the radar SNR and suppress SI signal as follows:
\begin{align}\label{eq: sub_opt2}
        \mathcal{OP}2:&\quad\mathbf{W}_{b}^{\rm RF} \triangleq \underset{\w_j\in\mathbb{F}_{\rm TX}}{\argmax}\quad \frac{\|(\mathbf{W}_{b}^{\rm RF})^{\rm H}\widehat{\H}_{\rm R} \mathbf{V}_{b}^{\rm RF}\|^2}{\|(\mathbf{W}_{b}^{\rm RF})^{\rm H}\widehat{\H}_{\rm b,b} \mathbf{V}_{b}^{\rm RF}\|^2}
\end{align}
Similar to $\mathcal{OP}1$, the solution of $\mathcal{OP}2$ requires a simple search through the available beam codebook $\mathbb{F}_{\rm RX}$. Given the analog SI cancellation taps $N\leq {N}_{b}^{\rm RF}{M}_{b}^{\rm RF}$ and the analog TX/RX beamformer at the BS, we follow a similar procedure to \cite{alexandropoulos2020full}, aiming to find the digital beamforming matrix $\mathbf{V}_{b}^{\rm BB}$ and the SI cancellation matrices maximizing the DL rate and suppressing the SI signal power below the RF saturation level of $\lambda_b$. The latter will ensure proper reception of the Radar target reflected signal. Our solution for the optimization problem \eqref{eq: optimization_eq} is summarized in Algorithm \ref{alg:the_opt}.

\begin{figure}[!tpb]
	\begin{center}
	\includegraphics[width=0.86\linewidth]{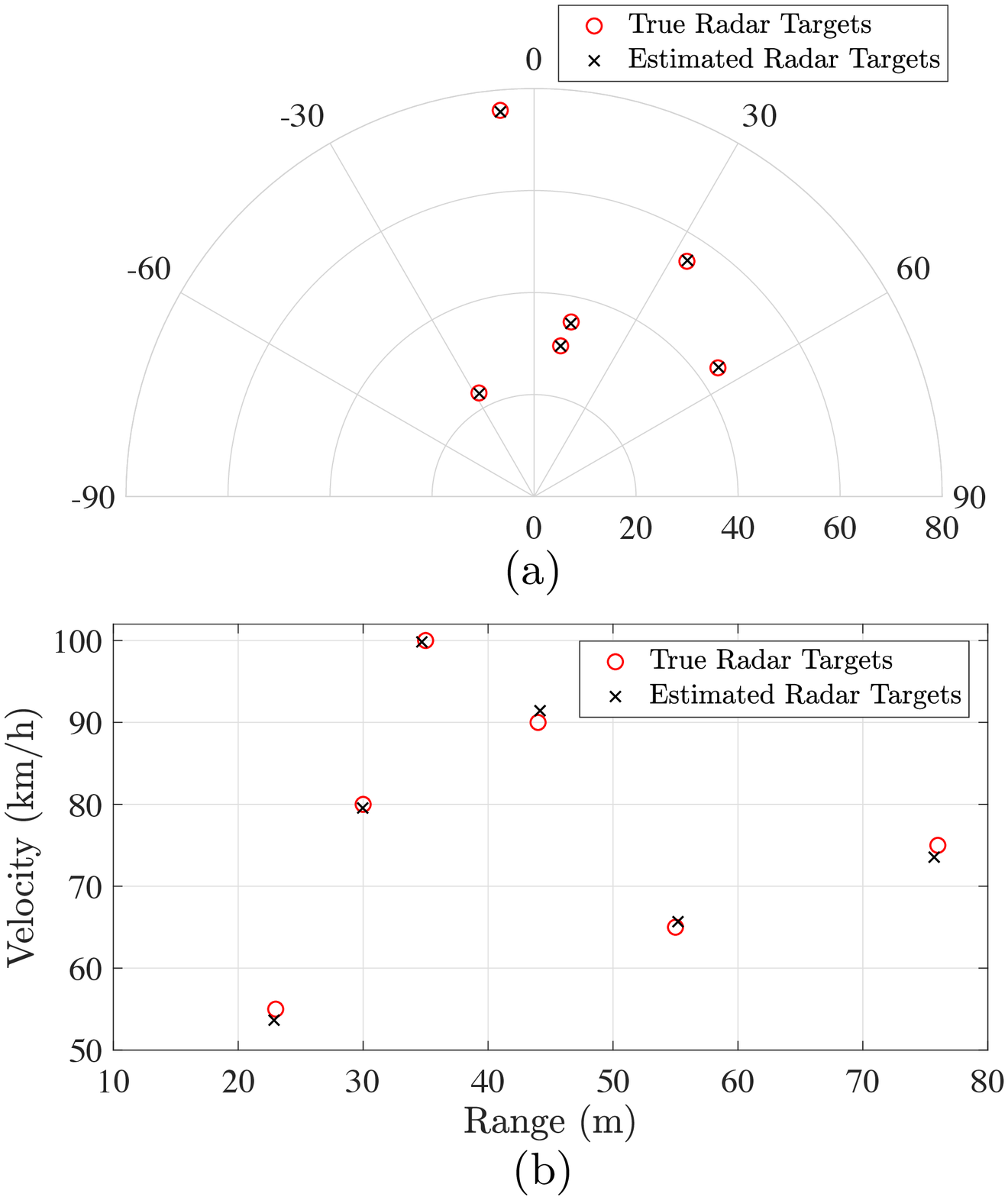}
	\caption{Sensing parameter estimation performance for $6$ radar targets with $N_{b}^{\rm RF} = M_{b}^{\rm RF} = 8, N_{b}^{\rm A} = N_{b}^{\rm A} = 16$  and transmit power of $30$dBm. (a)
	DoA and range estimation, (b) Range and relative velocity estimation.}\vspace{-0.2cm}
	\label{fig: Fig2}
	\end{center}
\end{figure}

\section{Numerical Results}
In this section, we present numerical results for the radar sensing and DL rate performance of the proposed FD massive MIMO ISAC system  operating at mmWave frequencies. 

\subsection{Simulation Parameters}\label{ssec: Sim_param}
We perform an extensive waveform simulation following the FD massive MIMO architecture illustrated in Fig.~\ref{fig: FD_ISAC} when operating at mmWave frequencies, where a $128\times 128$ FD massive MIMO node $b$ is communicating in the DL direction with $4$ antenna RX UE node $u$. The BS node $b$ employs $N_b^{\rm RF} = M_b^{\rm RF} = 8$ TX/RX RF chains with each of them connected to a ULA of $N_b^{\rm A} = M_b^{\rm A} = 16$ antenna elements via phase shifters.
The communication is performed using mmWave frequency of $28$GHz and a 5G NR OFDM waveform with $100$MHz BandWidth (BW) and $\Delta f = 120$KHz subcarrier spacing. According to the 5G NR specifications, we have $66$ Physical Resource Blocks (PRBs) resulting in $P=792$ active subcarriers and $Q=14$ OFDM symbols in each communication slot. Total OFDM symbol duration is defined as $T_s = 8.92\mu$s. We have considered a radio subframe of $1$ms for DL communication. 
The RX noise floors at all nodes were assumed to be $-90$dBm for $100$MHz BW OFDM signal. To this end, the RXs have an effective dynamic range of $60$dB provided by $14$-bit Analog-to-Digital Converters (ADC) for a Peak-to-Average-Power-Ratio (PAPR) of $10$ dB. Therefore, the residual SI power after analog SI cancellation at the input of each RX RF chain has to be below $-30$dBm to avoid signal saturation. The pathloss of the clustered DL channel is assumed to be $100$dB, whereas the SI channels are modeled as Rician fading channels with a $\kappa$-factor of $35$dB and pathloss $40$dB \cite{alexandropoulos2017joint}.
For the BS analog TX/RX beamformer, we consider a $5$-bit beam codebook based on the Discrete Fourier Transform (DFT) matrix.
We have used $1000$ independent Monte Carlo simulation runs to calculate the Radar sensing and DL rate performance.
\squeezeupan
\subsection{Radar Target Parameters}
We have considered $K=6$ radar targets randomly distributed in the sensing/communication environment with DoAs $\theta_k\in[-90^{\circ}\quad 90^{\circ}],\forall k$. For communication scatters, $L=2$ out of $K=6$ targets are chosen randomly. Each of the radar targets the range and relative velocity is selected randomly with a maximum range of $80$m and maximum velocity of $100$km/h.

\begin{figure}[!tpb]
	\begin{center}
	\includegraphics[width=0.84\linewidth]{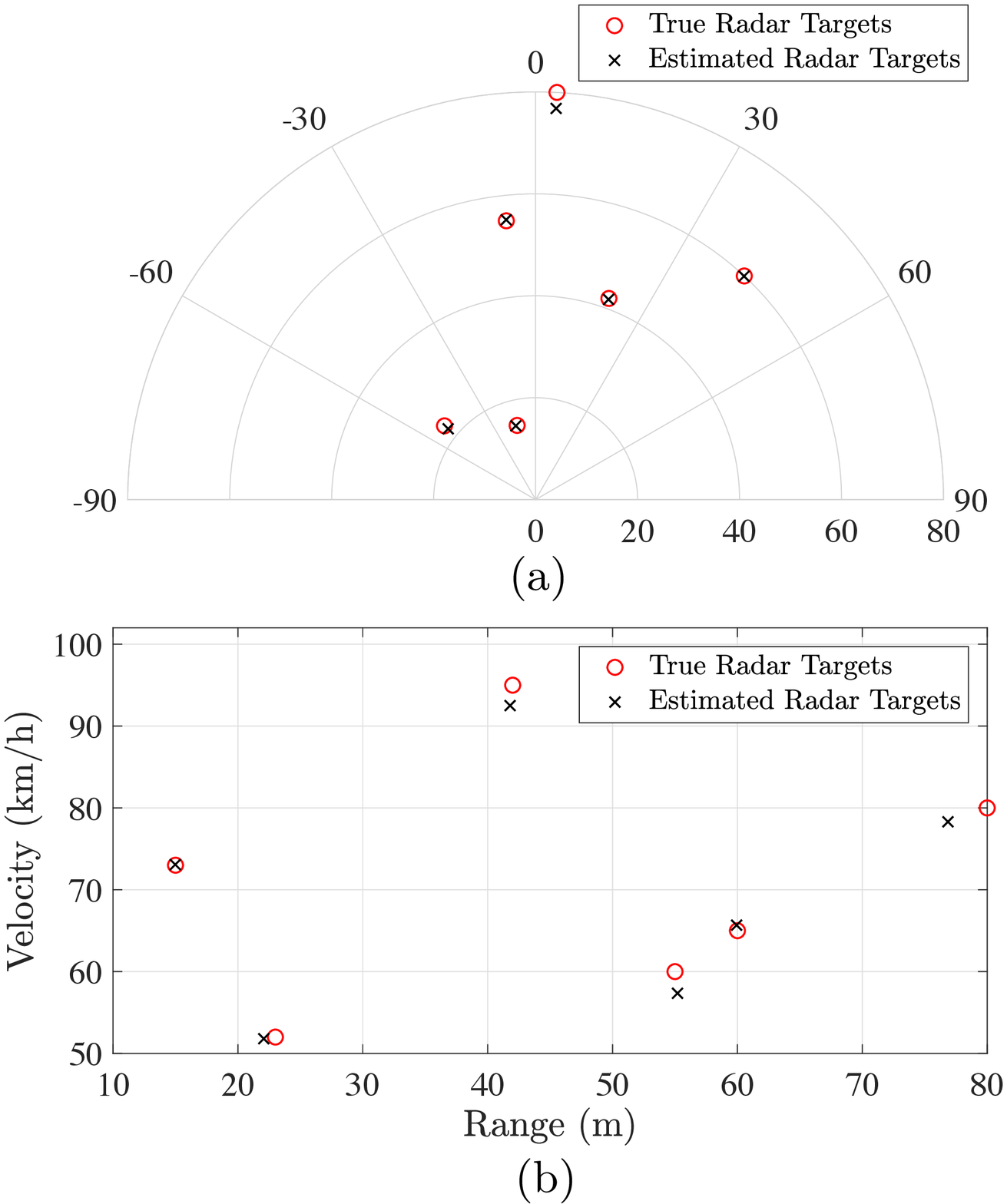}
	\caption{Sensing parameter estimation performance for $6$ radar targets with $N_{b}^{\rm RF} = M_{b}^{\rm RF} = 8, N_{b}^{\rm A} = N_{b}^{\rm A} = 16$  and DL transmit power of $10$dBm. (a)
	DoA and range estimation, (b) Range and relative velocity estimation.}\vspace{-0.2cm}
	\label{fig: Fig3}
	\end{center}
\end{figure}
\squeezeupan
\subsection{Radar Target Sensing performance}
In Fig.~\ref{fig: Fig2}, We have depicted the sensing performance of the proposed FD ISAC system with a $128\times 128$ massive MIMO node transmitting DL signal with a transmit power of $30$dBm. The DoA and range estimation is plotted in contrast to the true target parameters in Fig.~\ref{fig: Fig2}(a), where it is evident that the proposed FD ISAC system can detect all $6$ targets successfully with high precision. Even for really close target ($10^{\circ}$ and $12^{\circ}$) with only $2^{\circ}$ angle and less than $5$m range difference, the estimation performance is highly accurate. The superior sensing performance is provided by the proposed associated delay estimation approach with high-resolution MUSIC DoA estimation unlike previous FD ISAC work in \cite{liyanaarachchi2021joint}, where range estimation for such close targets was not possible. In Fig.~\ref{fig: Fig2}(b), the relative velocity is plotted with respect to the range estimation for $30$dBm DL transmit power. The figure shows that the proposed FD ISAC system is capable of estimating the relative velocity of all $6$ targets with less than $1.5\%$ estimation error.

In Fig.~\ref{fig: Fig3}, radar target sensing performance of the proposed FD ISAC system is presented for $10$dBm DL transmit power. It is evident from Fig.~\ref{fig: Fig3}(a) that the DoA estimation is almost accurate even at a low transmit power of $10$dBm. However, for the target at $3^{\circ}$, the estimated range is around $3$m away the actual value of $80$m. This is due to the low transmit power and higher path loss of the furthest target. In Fig.~\ref{fig: Fig3}(b), the relative velocity estimation is showcased, where the proposed approach achieved sensing performance with less than $5\%$ estimation error for a low transmit power of $10$dBm.

\begin{figure}[!tpb]
	\begin{center}
	\includegraphics[width=0.99\linewidth]{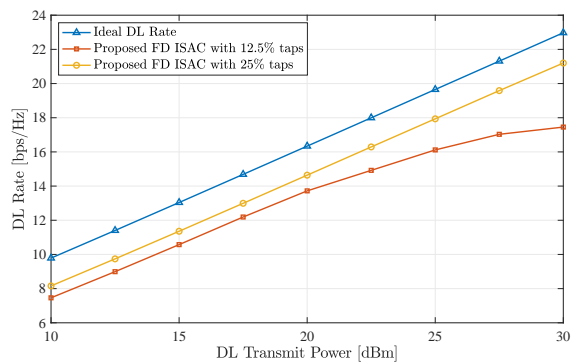}
	\caption{DL rate with respect to transmit power in dBm for the $128\times 128$ massive MIMO FD BS communicating with a $4$ antenna UE RX node. }\vspace{-0.2cm}
	\label{fig: DL_rate}
	\end{center}
\end{figure}
The DL rate performance of the proposed FD ISAC system with a $128\times 128$ massive MIMO BS node transmitting to a $4$ antenna UE RX node is depicted in Fig.~\ref{fig: DL_rate} with respect to transmit power. It is evident from the figure that the proposed FD ISAC approach is capable of providing DL rate very close ($1.5$bps/Hz) to the ideal rate in addition to the high radar sensing performance at the RX of BS node for $N=16$ ($25\%$) analog SI cancellation taps at the FD massive MIMO node. For even smaller hardware complexity ($12.5\%$) taps, the DL rate performance is comparable up to $20$ dB transmit power. However, as transmit power increases the impact of SI signal worsens, for DL transmit power of $30$dBm, the FD ISAC system is capable of providing $75\%$ of the ideal DL rate with only $N=8$ SI cancellation taps.

\squeezeup
\section{Conclusion}
In this paper, we presented an FD-based ISAC optimization framework, where an FD massive MIMO BS node is transmitting DL signals and concurrently performing radar target sensing utilizing the reflected signals. We devised a DoA, delay, and Doppler shift estimation algorithm for multiple radar target sensing considering hybrid A/D beamforming at the BS node. Adopting a limited complexity analog SI cancellation architecture, we presented a joint design of the A/D beamformer and SI cancellation that maximizes the DL rate together with the target sensing performance. Our performance results for a mmWave channel model demonstrated the high precision DoA, range, velocity estimation of multiple radar targets while providing maximized DL rate. In future work, the proposed FD ISAC approach will be considered for sensing and communication demanding practical applications, such as autonomous vehicles and flight control systems.
\squeezeupann
\section*{Acknowledgments}
\squeezeupann
This work was partially funded by the National Science Foundation CAREER award \#1620902.

\squeezeupann
\squeezeupann

\bibliographystyle{IEEEtran}
\bibliography{IEEEabrv,ms}

\end{document}